\documentclass[12pt,letterpaper]{article}  
\usepackage[includeheadfoot,
            marginratio={1:1,2:3}, 
            width=450pt, 
            height=688pt,]{geometry}

\usepackage{verbatim}
\usepackage{amsmath}
\usepackage{amsfonts}
\usepackage{amssymb}
\usepackage{mathtools}
\usepackage{physics}
\usepackage{graphicx}
\usepackage{tikz} 
\usetikzlibrary{decorations.pathreplacing,calc,calligraphy}
\usepackage{booktabs}
\usepackage{adjustbox}
\usepackage[utf8]{inputenc}
\usepackage{multirow}
\usepackage[splitrule]{footmisc}

\usepackage{empheq}
\usepackage{paralist}
\usepackage{cite}
\usepackage[hidelinks]{hyperref}
\usepackage{xcolor}
\usepackage{tensor}

\usepackage{nicefrac}

\newcommand{\nc}{\newcommand}
\nc{\lb}{\llbracket}
\nc{\rb}{\rrbracket}
\nc{\gl}{\llbracket}
\nc{\gr}{\rrbracket}
\nc{\del}{\partial}
\nc{\tri}{\hspace{-3.5pt}\vartriangle\hspace{-3.5pt}}
\nc{\blacktri}{\blacktriangle}

\nc{\eq}[1]{\begin{equation}
                     \begin{split} #1 \end{split}
                     \end{equation}}
\nc{\ul}{\underline}
\nc{\ov}{\overline}

\nc{\fa}{\hat}
\nc{\fb}{\MakeUppercase}
\nc{\fc}{\tilde}
\nc{\Lie}{{\cal L}} 
\nc{\lambdabar}{{\mkern0.75mu\mathchar '26\mkern -9.75mu\lambda}}

\allowdisplaybreaks[2]
\numberwithin{equation}{section}

\DeclareUnicodeCharacter{2212}{-}
\begin{document}

\vspace*{-1.5cm}
\begin{flushright}
  {\small
  MPP-2022-94\\
  }
\end{flushright}

\vspace{0.5cm}
\begin{center}
  {\LARGE
    The Dark Dimension in  a  Warped Throat 
} 
\vspace{0.2cm}

\end{center}

\vspace{0.15cm}
\begin{center}
Ralph Blumenhagen$^{1}$,
Max Brinkmann$^{2,3,4}$,
Andriana Makridou$^{1}$, \\[0.2cm]
\end{center}

\vspace{0.0cm}
\begin{center} 
{\footnotesize
\emph{
$^{1}$ 
Max-Planck-Institut f\"ur Physik (Werner-Heisenberg-Institut), \\ 
F\"ohringer Ring 6,  80805 M\"unchen, Germany } 
\\[0.1cm] 
\vspace{0.25cm} 
\emph{$^{2}$  Dipartimento di Fisica e Astronomia, Universit\`a di Bologna, \\ via Irnerio 46, 40126 Bologna, Italy}\\[0.1cm]
\vspace{0.25cm} 
\emph{$^{3}$  Dipartimento di Fisica e Astronomia, Universit\`a di Padova, \\via Marzolo 8, 35131 Padova, Italy}\\[0.1cm]
\vspace{0.25cm} 
\emph{$^{4}$  INFN, Sezione di Padova, \\via Marzolo 8, 35131 Padova, Italy}\\[0.1cm]
\vspace{0.3cm} 
}
\end{center} 

\vspace{0.3cm}

%%%%%%%%%%%%%%%%%%%%%%%%%%%%%%%%%%%%%%%%%%%%%%%
%%%%%%%%%%%%%%%%%%%%%%%%%%%%%%%%%%%%%%%%%%%%%%%
%%%%%%%%%%%%%%%%%%%%%%%%%%%%%%%%%%%%%%%%%%%%%%%
%%%%%%%%%%%%%%%%%%%%%%%%%%%%%%%%%%%%%%%%%%%%%%%
%%%%%%%%%%%%%%%%%%%%%%%%%%%%%%%%%%%%%%%%%%%%%%%
%%%%%%%%%%%%%%%%%%%%%%%%%%%%%%%%%%%%%%%%%%%%%%%
%%%%%%%%%%%%%%%%%%%%%%%%%%%%%%%%%%%%%%%%%%%%%%%
%%%%%%%%%%%%%%%%%%%%%%%%%%%%%%%%%%%%%%%%%%%%%%%

\begin{abstract}
  By combining swampland conjectures with observational
  data, it was suggested that our universe should lie
  in an asymptotic region of the quantum gravity landscape.
  The generalized distance conjecture for dS, the smallness of the
  cosmological constant and astrophysical 
  constraints led to a scenario with one mesoscopic large dimension of 
  size $\ell\sim \Lambda^{-1/4}\sim 10^{-6}$m.
  We point out that a strongly warped throat with
  its redshifted KK tower provides
  a natural string theoretic  mechanism that realizes the scaling
  $m\sim \Lambda^{\alpha}$ with the factor
  $\alpha=1/4$, the dark dimension being
  the one along the throat.
We point out    that in string theory it could  be  challenging
  to keep other KK towers heavy enough to avoid a conflict with
  astrophysical constraints  on the number of extra large dimensions.
\end{abstract}

\clearpage

%%%%%%%%%%%%%%%%%%%%%%%%%%%%%%%%%%%%%%%%%%%%%%%
%%%%%%%%%%%%%%%%%%%%%%%%%%%%%%%%%%%%%%%%%%%%%%%
%%%%%%%%%%%%%%%%%%%%%%%%%%%%%%%%%%%%%%%%%%%%%%%
%%%%%%%%%%%%%%%%%%%%%%%%%%%%%%%%%%%%%%%%%%%%%%%
%%%%%%%%%%%%%%%%%%%%%%%%%%%%%%%%%%%%%%%%%%%%%%%
%%%%%%%%%%%%%%%%%%%%%%%%%%%%%%%%%%%%%%%%%%%%%%%
%%%%%%%%%%%%%%%%%%%%%%%%%%%%%%%%%%%%%%%%%%%%%%%
%%%%%%%%%%%%%%%%%%%%%%%%%%%%%%%%%%%%%%%%%%%%%%%

\section{Introduction}
\label{sec:intro}

Since the recent swampland program (see
\cite{Palti:2019pca,vanBeest:2021lhn} for reviews)
postulates that not all self-consistent
quantum field theories admit a UV completion into a theory of quantum
gravity, the notion of naturalness has to be changed. What seems
natural from a purely low-energy point of view, could turn out to be in the
swampland after all. This new notion
has the potential to make quantum gravity and string theory
much more predictive than initially thought.
How such a logic can be made concrete was exemplified in the recent
work of Montero-Vafa-Valenzuela \cite{Montero:2022prj}. By combining swampland conjectures
with observational data, the authors  suggest that our universe should lie
in a specific corner of the quantum gravity landscape.

The starting point of \cite{Montero:2022prj} is
the assumption that our universe is located in an asymptotic region of field space
where the generalization of the AdS distance conjecture \cite{Lust:2019zwm}
to dS space applies. It states that for a cosmological constant $\Lambda$ approaching
zero, a tower of states becomes light obeying the  specific scaling behavior
\eq{
  m\sim |\Lambda|^\alpha\,.
  \label{ADC}
}  
Here a factor $\alpha\ge 1/2$ prevents scale separation between the internal dimensions and the radius of (A)dS space. It was
conjectured \cite{Lust:2019zwm} that for AdS this is always the case.
However, for dS the Higuchi bound \cite{Higuchi:1986py} already
requires that  $\alpha\le 1/2$. Taking one-loop
corrections into account, it was argued in \cite{Montero:2022prj} that
for four-dimensional dS the factor $\alpha$
should lie in the range ${1\over 4}\le \alpha\le {1\over 2}$.

Applying this to our universe with its observed tiny
cosmological constant
$\Lambda=10^{-122} M_{\rm pl}^4$ and taking bounds from
 deviations of Newton's  gravitational force law into account,
 revealed that only the value $\alpha=1/4$ can be consistent.
 Astrophysical constraints from the cooling/heating of neutron stars
 then led to a model of a single large extra  dimension, dubbed the
 dark dimension, and a corresponding tower of KK modes
 \eq{
   \label{darkdimrel}
   \Lambda^{1\over 4}=\lambda\, m_{\rm KK}
 }
 with $ 10^{-4}<\lambda<1$. 
 It was  further argued that the KK modes should include an excess
 of fermionic modes, leading to a kind of  sterile neutrino species.
 Moreover, the species scale \cite{Dvali:2007hz,Dvali:2007wp} related to this light tower
 of one-dimensional KK modes came out in the intermediate
 range $\Lambda_{\rm sp}\sim 10^9-10^{10}$GeV.
This is tantalizingly close to the scale of $10^{11}$GeV where
 due to the running of its self-coupling, the Higgs
potential is believed to develop an instability. 
Interpreting  the upper bound of  $10^{10}$GeV as the reason
for the sharp cutoff of the flux of ultra-high-energy cosmic
rays \cite{Anchordoqui:2022ejw},  led to the value  $\lambda =
 10^{-3}$. In \cite{Anchordoqui:2022txe}, it was argued that such a single mesoscopic
 direction also changes the production rate of primordial black holes
 such that they can provide a large fraction of the dark matter
 of our universe.  In  \cite{Gonzalo:2022jac} massive spin-2 KK modes along the dark dimension, dubbed dark gravitons,  were proposed as dark matter candidates,  leading to an explanation of the cosmological coincidence problem.

 The nice aspect of this derivation is that it only involves
 a generic swampland conjecture and observational data.
 However, consistency with one swampland conjecture
 does not guarantee that this Dark Dimension Scenario
 is really consistent with quantum gravity. Therefore,
 it is an important question whether it can be realized
 in a fully fledged theory of quantum gravity, like string theory.
 In this note we take a first modest step in settling  this question
 by pointing out that  a commonly used aspect 
 in string theory model building, a strongly warped throat, provides already
 a natural mechanism to get $\alpha=1/4$ in the generalized distance
 conjecture for dS while providing a lightest one-dimensional KK tower.
   
On the downside, we point out that generically
other KK modes and light states do not separate enough 
from the dark dimension to be compliant with 
astrophysical constraints.
This is exemplified in the LVS as a typical class of models.

 \section{Realization in a warped throat}
 \label{sec_two}

In string theory, realizations of AdS vacua are much better
understood than dS vacua. AdS vacua can for instance be realized via tree-level
flux compactifications and often give $\alpha=1/2$ in \eqref{ADC}.
It is fair to say that the question of scale separation for  AdS vacua
is not yet completely settled, as for instance the DGKT
vacua \cite{DeWolfe:2005uu} 
would be a candidate for an AdS
vacuum featuring scale separation.\footnote{ 
In an explicit example \cite{Blumenhagen:2019vgj}, it leads to a value of
$\alpha={7\over 18}$.} 
Consistent with the dS swampland conjecture
\cite{Obied:2018sgi} (see also \cite{Dvali:2014gua,Dvali:2018fqu}
for alternative arguments), it is fair to say that so far there does not
exist any generically accepted construction of a controlled
dS vacuum.

However, the KKLT construction \cite{Kachru:2003aw}  and the large volume
scenario (LVS) \cite{Balasubramanian:2005zx} are well studied candidates. 
These start
with an AdS minimum and then invoke an uplift mechanism to dS,
that is usually the contribution of an anti-D3-brane. To allow
a controlled balance of the AdS vacuum energy and the uplift
energy one puts the anti D3-brane in a strongly warped throat.
In this way, all energy contributions
localized in the throat get  redshifted
by the small warp factor.
By balancing the negative and positive contributions,
the former AdS minimum can be uplifted to a dS one.

Following this prescription, let us now assume that in a type IIB
orientifold set-up, by turning on 3-form fluxes on a Calabi-Yau manifold
and taking non-perturbative effects into account,
one has indeed arrived  at a not necessarily supersymmetric
AdS minimum with negative vacuum energy $V_{\rm AdS}$.
In order to be able to uplift, we also assume that the complex structure
moduli stabilization involves one modulus $Z$ that controls
the size of a 3-cycle, which for $Z\to 0$ approaches
a conifold singularity. This situation has been analyzed in detail
in e.g. \cite{Bena:2018fqc,Blumenhagen:2019qcg}  
and here we simply state and use some of their results.

Denoting the total volume modulus of the Calabi-Yau  as ${\cal
  V}$, for ${\cal V} |Z|^2\ll 1$ one is in the regime of strong
warping and the total geometry can be thought of as a long,
strongly warped Klebanov-Strassler (KS) throat \cite{Klebanov:2000hb}
of length $y_{\rm UV}$ glued to a bulk Calabi-Yau manifold.  Here
$y$ denotes the radial direction in the KS metric.
Following \cite{Douglas:2007tu}, the $N=1$ supersymmetric low energy effective action for the
two moduli $Z$ and ${\cal V}$ is described 
by a K\"ahler potential\footnote{It was shown in \cite{Lust:2022xoq} that
  this K\"ahler potential receives corrections when going off-shell.
  Note that here we are only interested in the behavior close to the minimum.} 
\eq{ 
\label{kaehlerpotb}
    K=-2\log({\cal V})   +  {2  c'  g_s M^2 \vert Z\vert^{2\over
      3}\over {\cal V}^{2\over 3}} + \dots\,
}
where the dots denote terms involving all the other complex structure
and K\"ahler moduli of the CY threefold. Moreover the string coupling
constant is related to the vacuum expectation values of the dilaton
$g_s=e^{\langle\phi\rangle}$, which is the real part of the
complex axio-dilaton $S=e^{-\phi}+iC_0$ field.
Computing the periods close to a conifold point $|Z|\ll 1$ and turning
on   R-R three-form flux $M$  along the A-cycle and
an NS-NS three-form flux $K$ along the B-cycle of the conifold
results in a superpotential of the form
\eq{
\label{superpotb}
     W=-{M\over 2\pi i} Z \log Z  +i  {K S} Z  +\ldots\,.
   }
where the dots again contain more flux induced tree-level
contributions involving  the complex structure moduli and
non-perturbative terms involving the K\"ahler moduli.
Then, the conifold modulus is stabilized at 
\eq{\label{Z_minimum}
Z\sim \exp\left(-{2\pi K\over  g_s M} \right)\,,
}
which self-consistently can be made exponentially small
by choosing appropriate fluxes.
For its mass one finds \cite{Blumenhagen:2019qcg}
\eq{
\label{massconi}
                m_Z\simeq  {1\over  (g_s M^2)^{1\over 2}} \left(
                  |Z|\over  {\cal V} \right)^{1\over 3}  \,.
              }
For an isotropic Calabi-Yau threefold one often considers the  bulk KK tower of
mass scale
\eq{
        M_{{\rm KK}}\sim {1\over \tau_b}\sim {1\over {\cal
            V}^{2\over 3}}\,.
        }
However, a compactification with a strongly warped throat is a 
highly non-isotropic situation, so that this does not necessarily 
reflect the lowest KK scale. 
In \cite{Blumenhagen:2019qcg} (see also \cite{Blumenhagen:2022dbo})
it was shown to first approximation analytically, and
confirmed numerically, that there exists a one-dimensional
tower of redshifted KK modes
that are mainly supported close to the tip of the KS throat.
Their masses scale in the same way as the mass of the conifold modulus
$Z$, i.e.\footnote{
More precisely, the numerical analysis \cite{Blumenhagen:2019qcg} indicated
that the localization of the KK modes close to the tip of the throat
makes the KK masses \eqref{masskkthroat} insensitive to the length of the throat 
$y_{\rm UV}$ beyond a critical length $y_{\rm UV} > y_{\rm UV}^{*}=O(10)$.
A further increase in throat length is not detected by the localized modes
and the scaling with $y_{\rm UV}$ stops.
} 
\eq{
  \label{masskkthroat}
       m_{{\rm KK}}\sim {1\over  (g_s M^2)^{1\over 2}\,y_{\rm UV}} \left(
                  |Z|\over  {\cal V} \right)^{1\over 3}  \,.
}
Note that while the warped modes contain a lower suppression
by the volume, the exponentially small value of $Z$
will easily make their mass scale much smaller than the bulk KK scale.
Control over  the warped effective action requires $g_s M^2\gg 1$ 
which additionally suppresses the warped KK scale.
The strongly warped throat can be  thought of as containing one
very long direction along the radial $y$-direction of the throat and thus being effectively 5-dimensional at
intermediate energy scales between  $m_{{\rm KK}}$ and $m^{(1)}_{{\rm
    KK}}$.
Here  $m^{(1)}_{{\rm KK}}$ denotes the mass scale of the second lightest tower of KK modes. This is not necessarily the bulk mass scale, as for instance
we expect the KK modes localized on the $S^3$ in the KS throat to also
become redshifted and thus be lighter than the bulk modes.
We will come back to this point at the beginning of section \ref{sec:IntermediateMasses}.

The uplift contribution to the scalar potential for an anti D3-brane placed
at the tip of the KS throat is given by \cite{Bena:2018fqc,Blumenhagen:2019qcg}  
\eq{
                V_{\rm up}\sim  {1\over  (g_s M^2)} \left(
                  |Z|\over  {\cal V} \right)^{4\over 3} \,.
}
We notice that parametrically (in the exponentially small quantity
$|Z|/{\cal V}$) we have the relation
\eq{\label{eq:KKthroat-Vup}
               m_{{\rm KK}}\sim    {1\over (g_s M^2)^{1\over
              4} y_{\rm UV}} \big|V_{\rm up}\big|^{1\over 4} 
}
between the warped KK mass and the uplift potential.
Choosing fluxes in \eqref{Z_minimum} 
such that $V_{\rm up}\sim |V_{\rm AdS}|$,
we get a relation between the value of $Z$ and the cosmological
constant of the AdS vacuum before the uplift
\eq{
                 { |Z|\over  {\cal V} }\sim   (g_s M^2)^{3\over 4}\,  \big|V_{\rm
                  AdS}\big|^{3\over 4}  \,.    
}

If the uplift really works and there is a meta-stable dS minimum,
the cosmological constant in the dS minimum  is then
given by $\Lambda=V_{\rm up}+V_{\rm AdS} \gtrsim 0$.
Note that  with respect to the exponentially small and therefore most
relevant parameter $|Z|/{\cal V}$, the final cosmological constant
$\Lambda$ scales in the same way as $V_{\rm up}$ and $|V_{\rm AdS}|$.

The usual landscape philosophy says  that playing with the warp factor
allows the cosmological constant to be tuned to hierarchically
smaller values.\footnote{Additional tuning could also arise from a
  landscape of initial AdS minima.}
However, one has to keep in mind that according to
\eqref{Z_minimum}  the VEV
of $Z$ is determined by quantized fluxes.
The number of fluxes one can use is expected to
be bound from above by tadpole cancellation conditions,\footnote{
To quantify the generic relative tuning available for $V_{\rm up}$ by choosing discrete fluxes in
\eqref{Z_minimum} we define  the relative minimal distance 
$\lambda'\sim {|Z_1|^{\nicefrac{4}{3}}-|Z_2|^{\nicefrac{4}{3}}\over |Z_1|^{\nicefrac{4}{3}}}$ 
for two values $|Z_1|>|Z_2|$.
Assuming that the tadpoles restrict the fluxes $M$ and $K$ to be
smaller than $N\gg 1$
one can derive the lower bound
\eq{\lambda'\sim \Big(1-e^{-{8\pi\over
    3g_s} {|K_2M_1-K_1 M_2|\over M_1 M_2}}\Big) >\Big(1-e^{-{8\pi\over
    3 g_s
    N^2}}\Big)\sim  {8\pi\over 3 g_s N^2}\,.\nonumber
}
For realistic values $g_s=1/10$ and $N=200$ one gets $\lambda'> 2\cdot
10^{-3}$. 
} the genuine
quantum gravity constraints in string theory.
Moreover, as shown \cite{Lust:2022xoq} the minimum should not move too far away
from its initial position.  This suggests that in quantum gravity the
actual tuning
one is allowed to do  in a controllable  manner is
limited. We include this tuning in our analysis  by writing
\eq{
                        \Lambda=\lambda'\, |V_{\rm up}|
}
with $\lambda'<1$.
Finally we use eq. \eqref{eq:KKthroat-Vup} to arrive at the relation
\eq{
                  \Lambda^{1\over 4}=    \Big[ (g_s M^2)^{1\over 4}y_{\rm UV}\lambda'\Big]\,    m_{{\rm KK}}\,.
}
We observe that this is precisely the relation \eqref{darkdimrel} for the dark
dimension scenario with $\lambda=(g_s M^2)^{1\over
  4}y_{\rm UV}\lambda'$. Imposing the bound $\lambda>10^{-4}$ from \cite{Montero:2022prj},
guaranteeing that a small value of $\lambda$
does not change the scaling too much, leads to
\eq{
       \lambda'>{10^{-4}\over (g_s M^2)^{1\over
  4}y_{\rm UV}}\,.
}
Consistent with what we just discussed, this restricts the amount
of ``fine tuning'' one can perform to get a small cosmological constant.

Note that if the former AdS vacuum
admits an uplift of this type, i.e. that fluxes can be found so that
the uplift condition $V_{\rm up}\sim |V_{\rm AdS}|$ holds and a meta-stable vacuum appears,
then it satisfies the scaling of the AdS distance conjecture with $\alpha=1/4$,
where the tower of states is given by KK modes in the long throat.
Thus, not only the uplifted dS but also the AdS vacuum is scale separated.

We conclude that an uplift by an anti D3-brane in a  strongly
warped throat generically leads to a tower of one-dimensional KK-modes
parametrically satisfying the (A)dS distance conjecture with $\alpha=1/4$.
In view of the unsettled control
issues raised recently for both the
KKLT scenario\cite{Gao:2020xqh,Lust:2022lfc,Blumenhagen:2022dbo} and
the LVS\cite{Junghans:2022exo,Gao:2022fdi}, one might wonder whether
one can draw any lesson from our analysis in case
this idea of initial AdS with subsequent uplifting
does not really work. Clearly, our result persists
as long as there exists a strongly warped KS throat and
the final (quasi) dS vacuum or even quintessence
potential is dominated by the energy scale in the strongly warped throat.

\section{Intermediate mass scales}
\label{sec:IntermediateMasses}

While this is quite a robust result, it is of course
just a single aspect of a whole string model. Putting aside
the not yet settled question whether dS vacua exist at all in string theory,
there are more mass scales in the game, like the other moduli masses
or other heavier KK towers that might still be in conflict with
astrophysical bounds. Their presence will also change
the estimate of the species scale, where gravity becomes
strongly coupled $\Lambda_{\rm sp}^{\rm grav}= M_{\rm
  pl}/\sqrt{N}=m_{\rm KK}^{1/3}M^{2/3}_{\rm pl}\sim 10^{9-10}$GeV. 
We have already indicated that beyond the lightest one-dimensional
tower of KK modes in the throat there might be more  towers
that are also  red-shifted and hence lighter than $\Lambda_{\rm sp}^{\rm grav}$.
Moreover, employing the emergence proposal  \cite{Heidenreich:2017sim, Grimm:2018ohb,Heidenreich:2018kpg}, it was
argued in \cite{Blumenhagen:2019qcg} that the effective theory in the throat determined by the
K\"ahler potential \eqref{kaehlerpotb}  and the superpotential
\eqref{superpotb} comes with its own cutoff
$\Lambda_{\rm sp}^{\rm throat}\sim (g_s M^2 y_{\rm UV})^{2\over 3} m_{\rm KK}  $ 
which is much lighter than $\Lambda_{\rm sp}^{\rm grav}$. 
Above this scale other non-perturbative
states appear that need to be included in the effective action.

Clearly, this is a very generic aspect of realizing the dark dimension
scenario via a warped throat.
Finally, let us also discuss the appearance
of other light moduli and KK towers for the case of  an uplifted large volume
scenario.

\subsubsection*{Example: uplifted LVS}

For the definition of the LVS and its moduli stabilization
scheme and the resulting mass scales we refer the reader to the
existing literature \cite{Balasubramanian:2005zx,Conlon:2005ki}.
Here we only need to know that there are two K\"ahler moduli, 
the volume modulus $\tau_b\simeq \mathcal{V}^{\frac23}$
and $\tau_s$, where the second is stabilized at small radius by a
non-perturbative effect, whereas the first is stabilized
perturbatively by an intricate balancing of three terms at
\eq{
  {\cal V}\sim \sqrt{\tau_s}\,  e^{a\tau_s}\,.
  }
The value of the cosmological constant in the non-super\-symmetric AdS
minimum scales like
\eq{
           V_{\rm AdS}\sim -{1\over \tau_s}\, e^{-3a\tau_s}\sim
           {1\over {\cal V}^{3}}\,.
}
The masses of the small and the large K\"ahler moduli scale as
\eq{
  m_{\tau_b}\sim {1\over {\cal V}^{3\over 2}}\,,\qquad
  m_{\tau_s}\sim {1\over {\cal V}}\,.
}
Recalling that the fluxes are chosen such that $V_{\rm up} \sim |V_{\rm AdS}|$,
we express the value of the conifold modulus in terms of the volume
\eq{
|Z| \sim   (g_s M^2)^{3\over 4}\,  {\mathcal{V}^{-\frac54}}\,.
} 
Then taking the scaling of the warped KK scale 
$m_{{\rm KK}}\sim V_{\rm up}^{1\over 4}\sim 1/{\cal V}^{3\over 4}$ 
and the (naive) bulk KK mass scale $M_{{\rm KK}}\sim 1/{\cal V}^{2\over 3}$ 
into account we get the following
hierarchy of mass scales
\eq{
        m_{\cal V} < m_{\tau_s}< m_{\rm
          KK} < M_{\rm KK}\,.
 }
Thus, we see that all K\"ahler moduli (and also almost all the complex
structure moduli) are lighter than the warped KK scale. In particular this
means that in  LVS the conifold modulus is actually the heaviest
complex structure modulus.
 Note that since ${\cal V}|Z|^2\sim {\cal
  V}^{-3/2}$ we are indeed in the regime of strong warping.
As expected,  the bulk KK modes are indeed heavier than the ones
arising in the throat. 
However, for their ratio one finds
\eq{
  { M_{{\rm KK}}\over  m_{{\rm KK}}}
 \sim {\cal V}^{1\over 12}\sim \Lambda^{-{1\over 36}}\sim 2\cdot 10^{3}
}
so that the corresponding length scale in  the bulk is only by a factor of
$10^{-3}$ smaller than the length scale of the 
throat.\footnote{We can be a bit more precise by taking into account that
  the background  is highly non-isotropic. In this case we better  approximate ${\rm
    Vol}=r_b^5\, r_t$ and define the bulk KK scale 
  as $M_{\rm KK}\sim {1\over r_b}$. In this way we find
  ${ M_{{\rm KK}}\over  m_{{\rm KK}}}
 \sim {\cal V}^{\nicefrac{1}{10}}\sim \Lambda^{-{\nicefrac{1}{30}}}\sim 10^{4}$.
}
This puts the uplifted LVS in tension with the astrophysical bounds
on  KK modes in more than one large extra dimension.
Moreover, this means that new physics appear below the species
scale computed (naively) via the throat KK modes.

\section{Conclusions}

In this note we pointed out that a common aspect of
string theoretic dS constructions naturally gives rise to the requirements
of realizing the dark dimension scenario.
We argued that under fairly  generic assumptions,  the non-isotropy
caused by the presence of a  strongly warped throat leads
precisely to the required exponent   $\alpha=1/4$ in the
(A)dS distance conjecture. The reason for this is simply that
the energy density of an anti D3-brane at the tip of the throat
and the mass scale of the one-dimensional tower of redshifted KK modes
localized deep in the throat satisfy $m_{\rm KK}\sim V_{\rm
  up}^{1\over 4}$. Using  the anti D3-brane as an uplift
from an AdS minimum to dS parametrically correlates the uplift energy scale
with the dS cosmological constant. The important numerical factor of
$\lambda \sim 10^{-1}-10^{-3}$ 
in the dark dimension scenario is then related to the flux dependent
and therefore restricted ``tuning'' of $\Lambda$ relative to $V_{\rm up}$.
Our result is expected to persist even under the milder assumptions that 
there exists a strongly KS throat in the geometry  and
that the final quasi dS energy is dominated by the energy scale
in the strongly warped throat.

On the downside, we pointed out that generically the warped
throat will support more  (towers of) light states
below the gravity cutoff scale. In this respect, we identified
more redshifted KK towers, non-perturbative states
beyond the cutoff of the effective theory in the warped
throat and, as  shown concretely for the LVS, other light bulk moduli
fields and KK towers. The appearance of  such modes will lower the gravity cutoff  
and will be in  conflict with the astrophysical constraints
on the size of extra dimensions.

%%%%%%%%%%%%%%%%%%%%%%%%%%%%%%%%%%%%%%%%%%%%%%%
%%%%%%%%%%%%%%%%%%%%%%%%%%%%%%%%%%%%%%%%%%%%%%%
%%%%%%%%%%%%%%%%%%%%%%%%%%%%%%%%%%%%%%%%%%%%%%%
%%%%%%%%%%%%%%%%%%%%%%%%%%%%%%%%%%%%%%%%%%%%%%%
%%%%%%%%%%%%%%%%%%%%%%%%%%%%%%%%%%%%%%%%%%%%%%%
%%%%%%%%%%%%%%%%%%%%%%%%%%%%%%%%%%%%%%%%%%%%%%%
%%%%%%%%%%%%%%%%%%%%%%%%%%%%%%%%%%%%%%%%%%%%%%%
%%%%%%%%%%%%%%%%%%%%%%%%%%%%%%%%%%%%%%%%%%%%%%%
\section{Acknowledgements}

We thank Dieter L\"ust  for useful comments.

\vspace{0.4cm}
%%%%%%%%%%%%%%%%%%%%%%%%%%%%%%%%%%%%%%%%%%%%%%%
%%%%%%%%%%%%%%%%%%%%%%%%%%%%%%%%%%%%%%%%%%%%%%%
%%%%%%%%%%%%%%%%%%%%%%%%%%%%%%%%%%%%%%%%%%%%%%%
%%%%%%%%%%%%%%%%%%%%%%%%%%%%%%%%%%%%%%%%%%%%%%%
%%%%%%%%%%%%%%%%%%%%%%%%%%%%%%%%%%%%%%%%%%%%%%%
%%%%%%%%%%%%%%%%%%%%%%%%%%%%%%%%%%%%%%%%%%%%%%%
%%%%%%%%%%%%%%%%%%%%%%%%%%%%%%%%%%%%%%%%%%%%%%%
%%%%%%%%%%%%%%%%%%%%%%%%%%%%%%%%%%%%%%%%%%%%%%%
%\newpage

\bibliography{references}  
\bibliographystyle{utphys}

%%%%%%%%%%%%%%%%%%%%%%%%%%%%%%%%%%%%%%%%%%%%%%%
%%%%%%%%%%%%%%%%%%%%%%%%%%%%%%%%%%%%%%%%%%%%%%%
%%%%%%%%%%%%%%%%%%%%%%%%%%%%%%%%%%%%%%%%%%%%%%%
%%%%%%%%%%%%%%%%%%%%%%%%%%%%%%%%%%%%%%%%%%%%%%%
%%%%%%%%%%%%%%%%%%%%%%%%%%%%%%%%%%%%%%%%%%%%%%%
%%%%%%%%%%%%%%%%%%%%%%%%%%%%%%%%%%%%%%%%%%%%%%%
%%%%%%%%%%%%%%%%%%%%%%%%%%%%%%%%%%%%%%%%%%%%%%%
%%%%%%%%%%%%%%%%%%%%%%%%%%%%%%%%%%%%%%%%%%%%%%%
\end{document}